\RequirePackage[l2tabu, orthodox]{nag}

\documentclass[a4paper,UKenglish]{lipics-v2016}

\usepackage[utf8]{inputenc}

\usepackage{amsmath}
\usepackage{amsfonts}
\usepackage{graphicx}

\usepackage{cite}

\title{Trainyard is NP-Hard}
\titlerunning{Trainyard is NP-Hard}

\author[1]{Matteo Almanza}
\author[2]{Stefano Leucci}
\author[2]{Alessandro Panconesi}
\authorrunning{M. Almanza, S. Leucci, and A. Panconesi}

\affil[1]{E-mail: \texttt{almanza.1597415@studenti.uniroma1.it}}
\affil[2]{Dipartimento di Informatica, ``Sapienza'' Università di Roma, Italy. \\ E-mail: \texttt{\{leucci,ale\}@di.uniroma1.it}}

\Copyright{Matteo Almanza, Stefano Leucci, and Alessandro Panconesi}


\renewcommand{\epsilon}{\varepsilon}

\usepackage{xspace}

\newcommand{\np}{\mbox{\textsf{NP}}\xspace}
\newcommand{\pspace}{\mbox{\textsf{PSPACE}}\xspace}

\newcommand{\ty}{\textsc{Trainyard}\xspace}
\newcommand{\msat}{\textsc{Min-Mon-SAT}\xspace}

\newcommand{\gsatisfy}{\texttt{Cross-Satisfy}\xspace}
\newcommand{\gignore}{\texttt{Cross-Ignore}\xspace}
\newcommand{\gterminus}{\texttt{Terminus}\xspace}
\newcommand{\gand}{\texttt{AND}\xspace}
\newcommand{\greplicator}{\texttt{Replicator}\xspace}
\newcommand{\goneway}{\texttt{One-Way}\xspace}
\newcommand{\gonetimepass}{\texttt{One-Time-Pass}\xspace}

\usepackage{placeins}
\usepackage{float}
\def\copyrightline{\relax}
\DOIPrefix{}
\begin{document}

\maketitle	

\begin{abstract}
	Recently, due to the widespread diffusion of smart-phones, mobile puzzle games have experienced a huge increase in their popularity.
	A successful puzzle has to be both captivating and challenging, and it has been suggested that this features are somehow related to their computational complexity \cite{Eppstein}. Indeed, many puzzle games --such as Mah-Jongg, Sokoban, Candy Crush, and 2048, to name a few-- are known to be \np-hard  \cite{CondonFLS97, culberson1999sokoban, GualaLN14, Mehta14a}.
	In this paper we consider Trainyard: a popular mobile puzzle game whose goal is to get colored trains from their initial stations to suitable destination stations. We prove that the problem of determining whether there exists a solution to a given Trainyard level is \np-hard. We also \href{http://trainyard.isnphard.com}{\color{blue}provide} an implementation of our hardness reduction.
\end{abstract}

\section{Introduction}

Trainyard, in the words of its author, is ``\emph{a grid-based logic puzzle game where the goal is to get each train from its initial station to a goal station. Every train starts out a certain colour, and most puzzles require the player to mix and merge trains together so that the correctly coloured trains end up at the right stations.}'' \cite{ToP}.

The game was conceived in 2009 and was first released for iPhones in 2010. In less than five months it climbed the Apple App Store charts becoming the most downloaded application in Italy and the United Kingdom, and the second most downloaded in the United States \cite{Story, Week}. It belongs to a class of games known as \emph{casual games}, video games that are targeted at a mass audience, have intuitive rules, and do not require a long-term time commitment to play.
The advent of smart-phones boosted the diffusion of casual games and it is estimated that number of mobile games will surpass 1.8 billions in 2017 \cite{CasualReport}. It has been suggested that the reason of this success is somehow related to their computational complexity, as David Eppstein famously said \cite{Eppstein}:
\begin{quote}
\itshape
If a game is in P, it becomes no fun once you learn ``the trick'' to perfect play, but hardness results imply that there is no such trick to learn: the game is inexhaustible. [...]

There is a curious relationship between computational difficulty and puzzle quality. To me, the best puzzles are NP-complete [...].
\end{quote}
Over the years several challenging puzzles have been shown to be at least \np-hard, e.g., Mah-Jongg \cite{CondonFLS97}, Fifteen-Puzzle \cite{RatnerW90}, Rush Hour \cite{FlakeB02}, Sokoban \cite{culberson1999sokoban}, Super Mario Bros and other classical Nintendo games \cite{AloupisDGV15}, Bejeweled, Candy Crush and similar match-three games \cite{GualaLN14}, 2048 \cite{Mehta14a}, just to cite a few. We refer the reader to \cite{KendallPS08} for a 2008 survey, although other puzzles have been proved to be \np-hard ever since, and to \cite{hearn2009games} for a general framework for showing \np- and \pspace-hardness puzzles. It is also known that games exhibiting certain mechanics are \np-hard \cite{Viglietta14}.

Next section briefly describes the game rules;  in Section~\ref{sec:results} we define our problem and state our main theorem, and in Section~\ref{sec:reduction} we describe the details of our hardness reduction.



\section{Game Mechanics}

\begin{figure}
	\centering
	\subcaptionbox{}{ \includegraphics[scale=0.3]{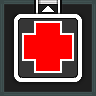} }
	\subcaptionbox{}{ \includegraphics[scale=0.3]{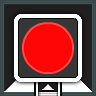} }
	\subcaptionbox{}{ \includegraphics[scale=0.3]{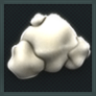} }
	\subcaptionbox{}{ \includegraphics[scale=0.3]{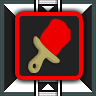} }
	\subcaptionbox{}{ \includegraphics[scale=0.3]{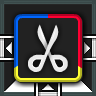} }
	\caption{Different kind of tiles: (a) departure station hosting one red train, (b) red arrival station with a capacity of one, (c) rock, (d) red painter, (e) splitter. Images courtesy of Matt Rix.}
	\label{fig:tiles}
\end{figure}

\begin{figure}[h]
	\centering
	\subcaptionbox[]{}{ \includegraphics[scale=0.3]{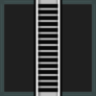} }
	\subcaptionbox[]{}{ \includegraphics[scale=0.3]{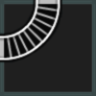} }
	\subcaptionbox[]{}{ \includegraphics[scale=0.3]{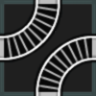} }
	\subcaptionbox[]{}{ \includegraphics[scale=0.3]{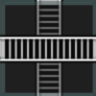} }
	\subcaptionbox[]{}{ \includegraphics[scale=0.3]{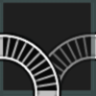} }
	\subcaptionbox[]{}{ \includegraphics[scale=0.3]{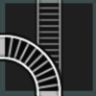} }
	\caption{Different kinds of rails. Each of the pieces can be rotated by $90$, $180$ or $270$ degrees.}
	\label{fig:rails}
\end{figure}

The game is played on board consisting of a rectangular grid divided into square cells. At the beginning, each cell of the board is either empty or it contains a special tile. There are several kind of tiles, the  most important being \emph{departure stations} and \emph{arrival stations}: the first ones host a number of trains while the latter are initially empty and have a maximum capacity, i.e, a number of trains then are able to hold (see Figure~\ref{fig:tiles}~(a) and (b)).\footnote{We will only use departure station hosting a single train and arrival stations with a capacity of one.}
The player can use empty cells to build different types of tracks. More precisely, the player can place any rail piece on each of the empty cells, where rail pieces can be either straight tracks, 90 degree turns, crossings, or switches (see Figure~\ref{fig:rails}). 
When the player is satisfied with his design, he can check his solution by simulating it.
The simulation proceeds in discrete time steps. At each step the board is updated as follows: all the departure stations that are not empty will output a train in one adjacent tile (depending on the initial orientation of the station), trains will move by one tile following the user's rails, and arrival stations that are not full will receive incoming trains if they are coming from the right tile (again, depending on the orientation of the arrival station).

If a train moves into an empty cell, in a cell where the rails have been misplaced, or out of the board, it crashes and the level is lost. Trains will also crash if they try to enter a special tile from the wrong direction (as we will discuss in the following) or if they try to enter in an arrival station that is at full capacity.
If the simulations reaches a state where all the trains have reached their arrival stations and each of these stations is at full capacity, without ever encountering a crash, then the player wins.

When two trains happen to be traveling on the same rail in the same direction, they \emph{merge} into a single train. If two trains \emph{touch} in any other way, they just proceed unobstructed, i.e., they pass through each other rather than crashing. 

\paragraph*{Colors}

To add more complexity, departure and arrival stations are colored: all the trains exiting a departure station will have its same color, while all the trains entering an arrival station must have a matching color or they will crash.
Trains can, however, change their color during their course due to special tiles and by touching or merging with other trains.
In our paper we only care of four colors: red, blue, purple and brown.
If two trains touch (resp. merge), their color will be modified (resp. the color of the resulting train will be chosen) according to the following rules. Let $A$ and $B$ be the color of the two trains and let $C$ be their new color (resp. the color of the resulting train). If $A$ and $B$ coincide, then we also have $C=A=B$. If one of $A$ and $B$ is red  and the other is blue, then $C$ will be purple. In all the remaining cases $C$ will be brown (as a consequence, if $A$ or $B$ are brown, then $C$ will be brown as well).

\paragraph*{Rails and switches}

The different kinds of rail pieces the user can place are shown in Figure~\ref{fig:rails}. The behavior of some of them is straightforward, while others deserve more attention. The user can rotate the pieces by $90$, $180$ or $270$ degrees before placing them; in the following we will refer to the orientation shown in figure.

Figure~\ref{fig:rails}~(d) shows two intersecting rails, these rails can be crossed in both the horizontal and vertical direction but turns are not allowed. Also note that two trains can cross the tile at the same time without crashing, as we already described.
Figure~\ref{fig:rails}~(e) shows a \emph{switch}. Notice that the rail turning to the left is highlighted, this means if a train comes from the bottom it will continue to the left and the switch will flip to the right; if another train arrives, it will turn to the right and the switch will flip again. If a single train comes from left or right, it will proceed towards the bottom regardless of the current state of the switch, but this will still cause the switch to flip.
If two trains come from the sides at the same time, they will merge into a single train and the switch will flip.
Figure~\ref{fig:rails}~(e) shows another type of switch, consisting of a straight track and a left turn; here similar rules to the ones we just described apply. When placing the rails, the player is able to choose the initial state of the switches. 

Finally, we note that if two trains are on different rails of the Figure~\ref{fig:rails}~(c), they will not touch and hence they won't cause their colors to mix.

\paragraph*{Special tiles}

Apart from the departure and arrival stations, that we already described (see Figure~\ref{fig:tiles}~(a) and (b)), there are also three other special tiles: the \emph{rock}, the \emph{painter} and the \emph{splitter}.
A rock is just a tile that causes any entering train to crash, effectively removing one position available for the player to place a rail piece (see Figure~\ref{fig:tiles}~(c)). 

A painter (shown in Figure~\ref{fig:tiles}~(d)) always has a color $C$ that is either red or blue, and will change the color of any incoming train to $C$. It has two inputs/outputs on opposite sides and can be traversed in both directions (a train entering from a side will exit from the opposite one). Any train trying to enter a painter from one of the other two sides will crash.

The splitter is more involved: it has only one input, marked in yellow, and two outputs in the sides adjacent to the input. One of the outputs (clockwise w.r.t.\ the input) is marked in red and the other one is marked in blue (see Figure~\ref{fig:tiles}~(e)). 
Whenever a train enters the splitter from the input, two trains will exit from the two outputs in the next time step (and the original train vanishes). The colors of these two new trains depend on the color $C$ of the input train: if $C$ is red, blue, or brown then the two new trains will also be colored $C$. If $C$ is purple, then the train exiting from the red-marked output will be red, and other train (exiting from the blue-marked output) will be blue. Any train trying to enter a splitter from a side different from its input side will crash.

\begin{figure}[t]
	\centering
	\subcaptionbox[]{}{ \includegraphics[scale=0.40]{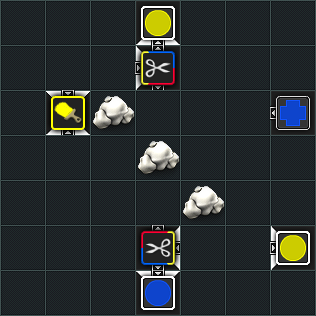} }
	\subcaptionbox[]{}{ \includegraphics[scale=0.40]{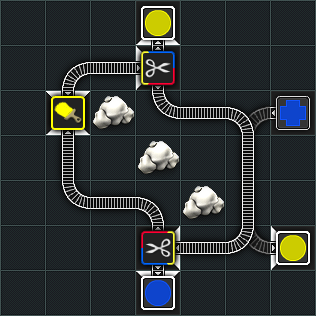} }
	\caption{A level of Trainyard (left) and a possible solution (right).}
	\label{fig:example}
\end{figure}

\section{Our Results}
\label{sec:results}

In this paper we consider the problem of deciding whether a given level of Trainyard admits a solution.
More precisely, we define \ty as the following decision problem: given a rectangular board and an initial placement of the tiles shown in Figure~\ref{fig:tiles}, is there a way to place the rails pieces shown in Figure~\ref{fig:rails} on (a subset of) the empty cells so that the simulation will reach a state where: (i) there are no trains left in the departure stations, (ii) there are no moving trains, (iii) no train crash has happened, and (iv) all the arrival station received a number of trains matching their capacity?

In the following section we will design a polynomial time reduction from a variant of the boolean satisfiability problem to \ty, hence proving the following:  
\begin{theorem}
	\label{thm:ty_is_np_hard}
	\ty is \np-hard.
\end{theorem}
We also provide an actual implementation of our reduction which can be found at \linebreak \url{http://trainyard.isnphard.com}.

We note that we do not actually know if \ty lies in \np, i.e., we do not know whether, given a level and a corresponding design for the rails, it is possible to check, in polynomial time, that the solution is indeed correct.
The trivial simulation strategy fails as it is possible to create solutions that require an exponential number of time steps for the simulation to stop. On the other hand, \ty is clearly in \pspace as the simulation algorithm only needs to keep track of the current state of the board. This requires polynomial space since, due to the merge rule, the number of moving trains cannot be asymptotically larger than the size of the board itself.
We regard the problem of establishing whether $\ty \in \np$ as an interesting --and fun-- challenge.

\section{Our Reduction}
\label{sec:reduction}

We prove the \np-hardness of \ty by showing a polynomial reduction from (the decision version of) \emph{Minimum Monotone Boolean Satisfiability Problem} (\msat for short). In \msat we are given (i) a CNF formula $\phi$ of $n$ variables $x_1, \dots, x_n$ and $m$ clauses $C_1, \dots, C_m$ such that each clause contains only positive literals, and (ii) an integer $k$. The goal is to decide whether there exists a truth assignment for the variables that satisfies $\phi$ and sets at most $k$ variables to true.
This problem is easily shown to be \np-hard as there is a straightforward reduction from the decision version of \emph{Minimum Dominating Set}: for each vertex $u$ of the input graph we add a variable $x_u$ and a clause $\bigvee_{v \in N[v]} x_v$, where $N[v]$ denotes the closed neighborhood of $v$. Clearly, there exists a dominating set of size $k$ iff the formula can be satisfied by setting at most $k$ variables to true.

In the following subsections we first give an high-level picture of our reduction and then we move to the description of the gadgets we use. 

\subsection{Overview}

\begin{figure}[t]
	\centering
	\includegraphics[width=0.80\textwidth]{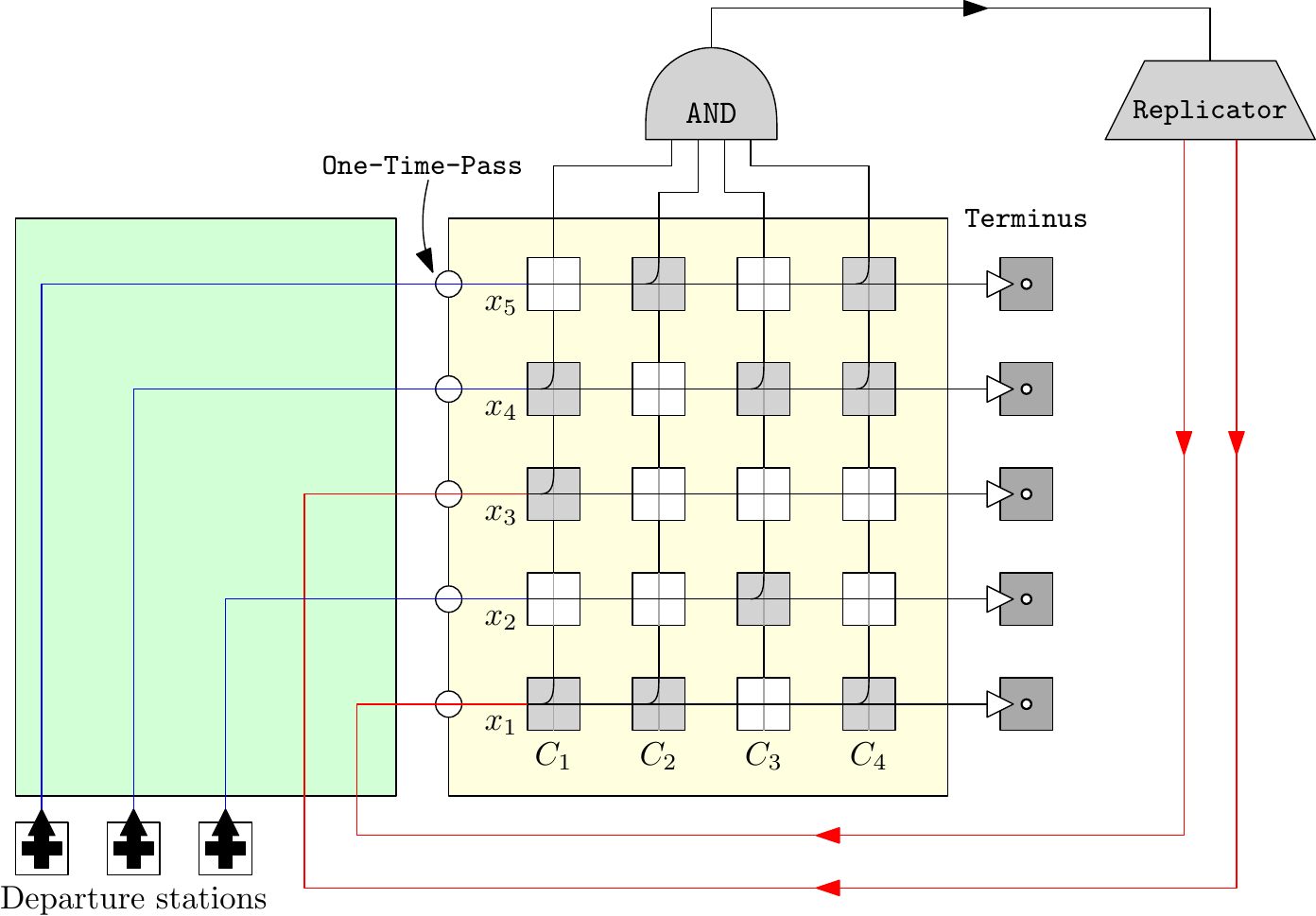}
	\caption{A high-level picture of our reduction for an instance of $\msat$ having $5$ variables, $4$ clauses, and $k=3$. The corresponding formula is $(x_1 \vee x_3 \vee x_4) \wedge (x_1 \vee x_5) \wedge (x_2 \vee x_4) \wedge (x_1 \vee x_4 \vee x_5)$. \gignore gadgets are shown in white while \gsatisfy gadgets are in gray.}
	\label{fig:overview}
\end{figure}

The overview of our reduction is shown in Figure~\ref{fig:overview}. In the bottom left we have a $k$ \emph{departure stations} --one per variable than can be set to true-- each containing a single train.
The \emph{clause area}, shown in yellow, encodes the formula $\phi$ using a $n \times m$ matrix of gadgets: the gadget on the $i$-th row\footnote{We are counting rows from bottom to top.} and $j$-th column will be one of two different types that we call \gsatisfy and \gignore. More precisely, we use a \gsatisfy gadget whenever $x_i$ is contained in $C_j$ and a \gignore gadget otherwise.

To describe how our reduction works, let us look at the case where the formula is satisfiable using $k$ true variables. Consider, e.g., the instance of Figure~\ref{fig:overview} and the satisfying assignment $x_2 = x_4 = x_5 = \textup{true}$, $x_1=x_3=\textup{false}$. Here the rails are designed so that the $k$ trains exiting from the departure stations will traverse the rows of the matrix corresponding to the variables set to true.
When a train enters a \gignore or \gsatisfy gadget from the left it will proceed to the right, thus entering the next gadget on the same row. After a train finishes traversing a row (i.e, when it exits from the right side of last the gadget on the row), it is collected by a dedicated \gterminus gadget containing an arrival station.
Moreover, when a train enters a \gsatisfy gadget from the left, it is possible to create a copy of it by suitably placing the rails. If this copy is created, the new train will necessarily exit from the top of the gadget.  Intuitively this means satisfying the clause corresponding to the column using the variable corresponding to the row.
Since we started with a satisfying assignment, it is possible to duplicate a train for each column. These trains will move from bottom to top: each time a train enters a \gignore or \gsatisfy gadget from the bottom it will only be allowed to exit from the top.
Eventually, all these trains will exit the clause area and they will enter the \gand gadget: this gadget allows a train to exit form the top iff all $m$ trains are entering from the bottom, i.e., iff all the clauses have been satisfied. 

Notice that, at this point, we still have to bring a train to the \gterminus gadgets of the rows corresponding to false variables, e.g., $x_1$ and $x_3$. Moreover, due to their actual implementation, the gadgets on these rows also need to be traversed by a train going in the left-right direction.
In order to successfully complete the level, we make $n-k$ copies of the single train exiting from the \gand gadget using a \greplicator gadget, and we feed them into such rows.
To guarantee that two or more trains can not enter the same row of the matrix, we a suitable \gonetimepass gadget which is placed at the beginning of each row.

Now consider the case where the $k$ departing trains traverse a set of rows whose corresponding variables do not satisfy the formula, e.g., $x_2, x_3$ and $x_4$. In this case it will not be possible to both (i) satisfy the \gterminus gadgets of these rows and (ii) allow a train for each column to reach the $\gand$ gadget.  As a consequence, if the formula is not satisfiable, every possible assignment will necessarily result in a loss.

It is to be noted that, although the player can place the rails in any empty tile of the board, our construction is such that the layout of the rails is essentially forced, except for the green area --which encodes the truth assignment-- and for the \gsatisfy gadgets where rails can be placed in two different ways, as we will discuss in the following.
 
\subsection{Handling Parity Issues}

Consider two train stations with one train each are placed next to each other, according to the mechanics of the game, the two exiting trains will never be able to merge. This can be easily seen by considering the \emph{parity} of the trains: if a train occupies the $i$-th row and the $j$-th column of the board, then its parity is $(i+j) \bmod 2$.
As the two stations are adjacent, the initial parity of the two trains will differ. Now, according to the mechanics of the game, each train moves by exactly one cell per time step, hence flipping its parity. We refer to the initial parity of a train as its \emph{phase}.

Actually, it is possible to show that two trains can merge iff they have the same phase (also notice that trains exiting from a splitter will have the same phase of the entering train).
In order to avoid being distracted by parity issues, in the following we will assume that all trains have the same phase. This can be easily achieved by restricting the placement of the departure stations to tiles in a checkerboard pattern.

\subsection{Description of the Gadgets}

Here we describe how to the gadgets used in our reduction can be implemented and we argue on their correctness.

\subsubsection{Rails and Lanes}

\begin{figure}[h]
	\centering
	\includegraphics[scale=0.4]{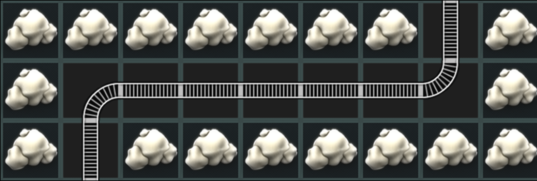}
	\caption{A lane and the corresponding design of rails.}
	\label{fig:lane}
\end{figure}

In our reduction we will need to move the trains from one gadget to another. However, as we cannot directly place rails in our instance, we need to force the player to build the correct railways between gadgets. This is easily done by creating a \emph{lane} of rocks, leaving a single empty tile in-between so that there is only one way for the player to design the rails without causing a train to crash. An example of a lane where rails have already been placed is shown in Figure~\ref{fig:lane}.

\subsubsection{Terminus Gadget}

\begin{figure}[!h]
	\centering
	\includegraphics[scale=0.4]{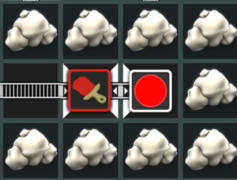}
	\caption{Implementation of the \gterminus gadget.}
	\label{fig:terminus}
\end{figure}

The \gterminus gadget is just an arrival station preceded by a painter to ensure that the incoming train will be of the correct color, as it is shown in Figure~\ref{fig:terminus}.

\subsubsection{One-Way Gadget}

\begin{figure}[!b]
	\centering
	\includegraphics[scale=0.4]{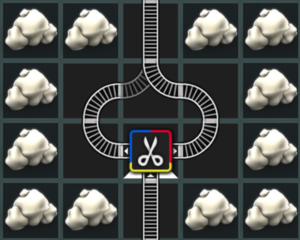}
	\caption{Implementation of the \goneway gadget.}
	\label{fig:oneway}
\end{figure}

The \goneway gadget has one input and one output on the opposite side. As the name suggests, it can only be traversed in the input-output direction. Any attempt to traverse the gadget in the opposite direction will cause a train to crash and hence the player to lose the level. 
Its implementation is shown in Figure~\ref{fig:oneway} and it consist of a single splitter with suitable spacing. Notice that there is only one way to design the rails for this gadget that does not cause an incoming train to crash. This gadget will be useful as a component in our other gadgets.

\subsubsection{One-Time-Pass Gadget}

\begin{figure}[!ht]
	\centering
	\includegraphics[scale=0.4]{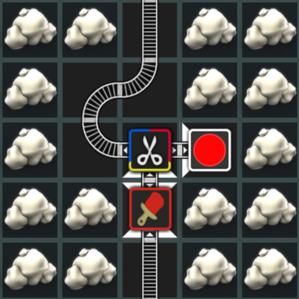}
	\caption{Implementation of the \gonetimepass gadget.}
	\label{fig:onetimepass}
\end{figure}

This gadget is similar to the \goneway gadget but it has the additional constraint that it must only be traversed exactly once in the whole level. Its implementation is a straightforward modification of the \goneway gadget and its shown in Figure~\ref{fig:onetimepass}.

\subsubsection{Cross-Ignore Gadget}

\begin{figure}[!ht]
	\centering
	\includegraphics[scale=0.4]{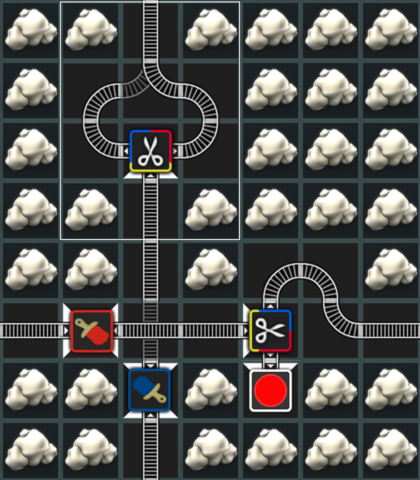}
	\caption{Implementation of the \gignore gadget. The \goneway subgadget is highlighted.}
	\label{fig:ignore}
\end{figure}

The \gignore gadget is used in the matrix in the clause area each time a variable $x_i$ does not appear in a clause $C_j$.
The only ways of placing the rails in this gadget without losing the level ensures that a train can exit from the right (resp. top) only if a train is entering from the left (resp. bottom). The gadget, along with such a design of the rails, is shown in Figure~\ref{fig:ignore}.
Notice that the train coming from the left will always be colored red while the one coming form the bottom will be colored blue.
To show the correctness of the gadget we only need to argue on rail piece placed on the tile adjacent to the two painters, since the rest of the rail design is forced. 
If the rails in that tile cross, as in figure, then it is clear that if only one train arrives the gadget works as expected. If two trains arrive, one from the left and one from the bottom, then either they touch or they do not. If they do not touch, then the red train is split so that one copy can continue to the right while the other goes into the arrival station. If they touch, then they will both become purple. This does not cause any problems since the train going to the right will be split into one blue and one red train: the red train goes into the arrival station and the blue train exits from the top of the gadget.

If the rails on the tile are placed in any way that causes that the left train to go up, then the arrival station will always remain empty (as the blue train cannot be sent there), thus losing the level. Otherwise, if the rails merge and continue to the right, then the level will be lost unless \emph{both} trains come at the same time. In that case the arrival station will receive a train and another train will exit from the right, but no train will exit from the top. This case, however, does not cause any problems since it corresponds to ``forgetting'' that $C_j$ has already been satisfied by some variable in $x_1, \dots, x_{i-1}$, which is never convenient.

Finally, notice that no train can come from the right (due to the final splitter) or from the top (due to the \goneway subgadget).

\subsubsection{Cross-Satisfy Gadget}
\begin{figure}[!b]
	\centering
	\subcaptionbox[]{}{ \includegraphics[width=0.45\textwidth]{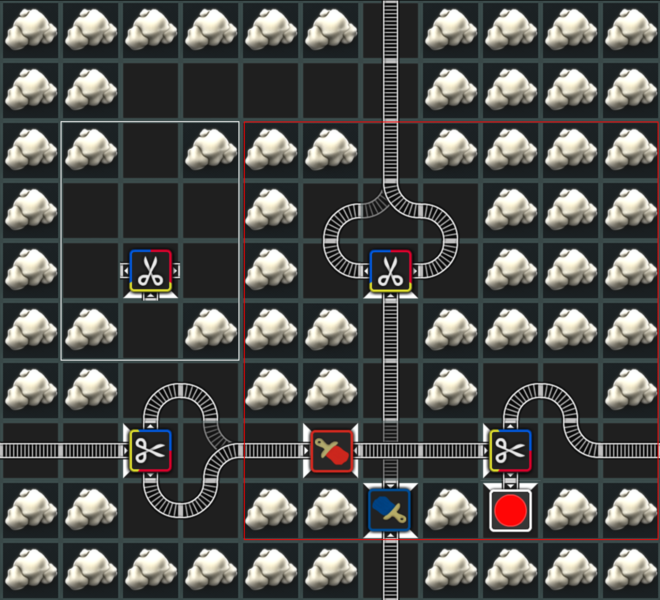} }
	\subcaptionbox[]{}{ \includegraphics[width=0.45\textwidth]{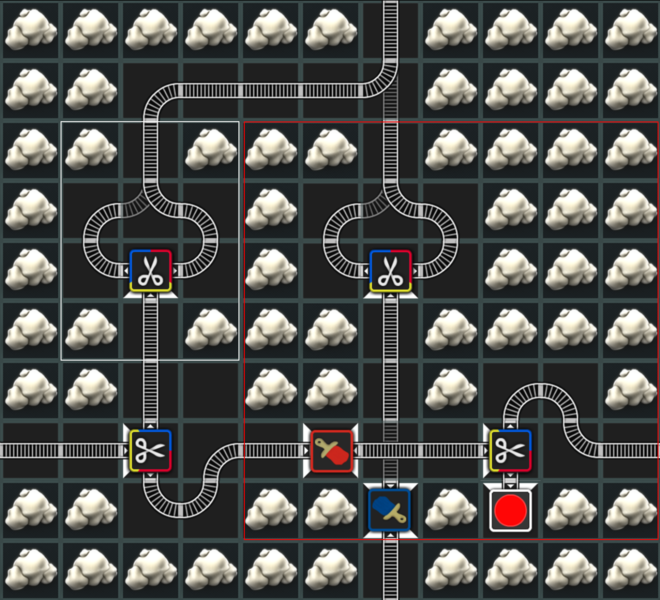} }
	\caption{Implementation of the \gsatisfy gadget with two possible rail designs. The \goneway and \gignore subgadgets are highlighted in white and red, respectively.}
	\label{fig:satisfy}
\end{figure} 

The \gsatisfy gadget is used in the matrix in the clause area each time a variable $x_i$ appears in a clause $C_j$.
It works similarly to the \gignore gadget but the player also has the option to place the rails so that a train is able to exit from the top when the left train reaches the gadget. This encodes the fact that the clause $C_j$ has been satisfied using variable $x_i$. %
The implementation is shown in Figure~\ref{fig:satisfy} where a \goneway a \gignore subgadgets are highlighted in white and red, respectively.
Apart from the \gignore subgadget that we already discussed, the only sensible rail designs that do not make the player lose the level, are those shown in Figure~\ref{fig:satisfy}~(a) and (b). The first one acts exactly as a \gignore gadget: the train entering from the left is just split and rejoined.
In the second design, the train is also split but now a copy continues towards the top while the other serves as an input for the \gignore gadget. The two rails going to the top are then joined together in a single rail which is the output of the gadget (here the two \goneway gadgets ensure that no train can go back into the gadget). 
Notice that, when the left and the bottom train both enter the gadget, the second design can cause two trains to exit from the top and one to reach the \gignore subgadget. 
However, doing this never helps in completing the level and the first design can be chosen instead.

\subsubsection{AND Gadget}

The \gand gadget takes a number of train as inputs from the bottom. If all these trains eventually reach the gadget, then the rails can be designed so that a single train will reach the top. On the converse, if at least one train is missing, then there is no way of placing the rails to make a train exit the gadget.
Here we describe how such a gadget with two inputs is implemented. In order to extend the construction to an arbitrary number of inputs trains it suffices to chain together multiple copies of this gadget.

\begin{figure}[!b]
	\centering
	\includegraphics[scale=0.35]{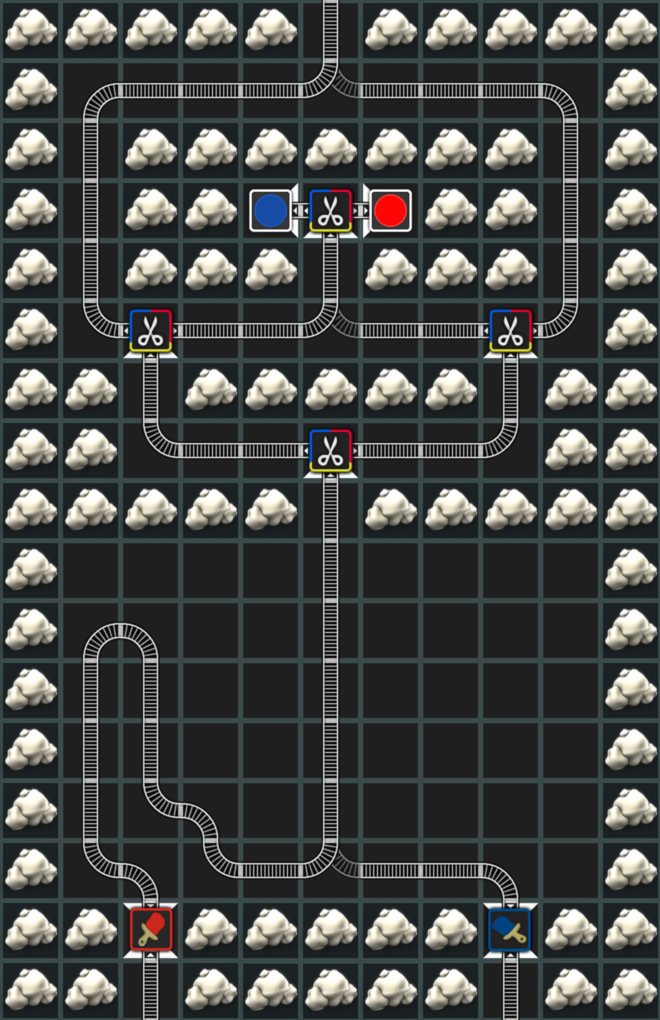}
	\caption{Implementation of the \gand gadget.}
	\label{fig:and}
\end{figure}

The implementation is shown in Figure~\ref{fig:and} where two distinct areas can be seen. The bottom one is a \emph{buffer} area: here 
there is some empty space where rails can be placed so that the incoming trains (which are colored in read and blue by the painters) can be merged into a single purple train.
This purple train can then proceed to the upper area. In this area there is only one possible rail design that will not result in a train crash, namely the one shown in Figure~\ref{fig:and}. It is easy to see that if a single purple train reaches the upper area, this correctly causes two trains (one red and one blue) to reach the arrival stations and one to exit from the top of the gadget. Any other number of trains or any single non-purple train will result in a crash.

\subsubsection{Replicator Gadget}

\begin{figure}[ht]
	\centering
	\includegraphics[scale=0.4]{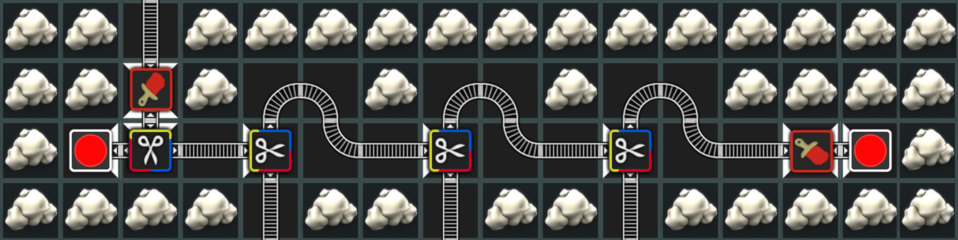}
	\caption{Implementation of a \greplicator gadget with three outputs.}
	\label{fig:replicator}
\end{figure}

This gadget takes a train from the top as an input and creates a number of copies of it, so that $n-k$ trains will exit from the bottom.
Intuitively, in a yes instance, these trains should enter the rows corresponding to the negated variables.
The implementation of a \greplicator with three outputs is shown in Figure~\ref{fig:replicator}. The construction can be adapted in a straightforward way in order to create as many copies of the incoming train as necessary.

\subsection{Final Remarks}

Combining together all the above gadgets, as indicated by Figure~\ref{fig:overview}, allows us to build the instance of \ty corresponding to the instance of \msat. We can now sketch the proof of Theorem~\ref{thm:ty_is_np_hard}, whose correctness follows from the correct operation of the single gadgets.

\begin{proof}[Sketch of the proof of Theorem~\ref{thm:ty_is_np_hard}]
If there is a truth assignment for \msat, then the $k$ trains exiting from the departure stations can be fed into the rows of the clause matrix corresponding to variables set to true.
For each clause $C_j$ let $r_j$ be the index of the first variable that satisfies $C_j$, i.e., $r_j = \min\{1 \le i \le n : x_i = \mbox{true} \wedge x_i \in C_j\}$. By construction, the $j$-th column of the matrix will have  a \gsatisfy gadget on the $r_j$-th row. This means that we can design the rails in the \gsatisfy gadgets of coordinates $(r_j, j)$ so that a train will exit from the top. The rails for all the other gadgets in the matrix  can be designed so that incoming trains just cross to the opposite sides. This allows exactly one train per column to reach the \gand gadget, and $n-k$ trains to exit the \greplicator. These trains are fed into the $n-k$ remaining rows to win the level.

On the other hand, if we have a solution for the instance of \ty, then this means that exactly one train is entering in each row of the matrix (since they are all preceded by a \gonetimepass gadget). As there are only $k$ departure stations, and the only way to make more trains reach the green area is by using the \greplicator gadget, this implies that at least $k$ trains reached the \gand gadget (one train per column). This, in turn, implies the existence of at least one \gsatisfy gadget per column where a train is entering from the left. Hence, this \gsatisfy gadget must necessarily be placed on a row traversed by a train coming from a departure station.
This means that we can find a satisfying truth assignment for the instance of \msat by setting to true the variables corresponding to the rows where the $k$ departing trains are entering.
\end{proof}

A solved instance of \ty corresponding to the formula $(x_1 \vee x_2) \wedge (x_2 \vee x_3)$ can be seen in Figure~\ref{fig:instance} (which is moved to the appendix due to space limitations). Moreover at \url{http://trainyard.isnphard.com} it is possible to generate instances of \ty corresponding to arbitrary (small) \msat formulas and even simulate the possible solutions.

\subparagraph*{Acknowledgements}

The authors wish to thank Erik D. Demaine and Rupak Majumdar for insightful and pleasant discussions and email exchanges.

\bibliographystyle{plainurl}
\bibliography{trainyard}

\clearpage
\appendix

\section{An Instance of Trainyard Arising from Our Reduction}

\begin{figure}[H]
	\centering
	\includegraphics[width=\textwidth]{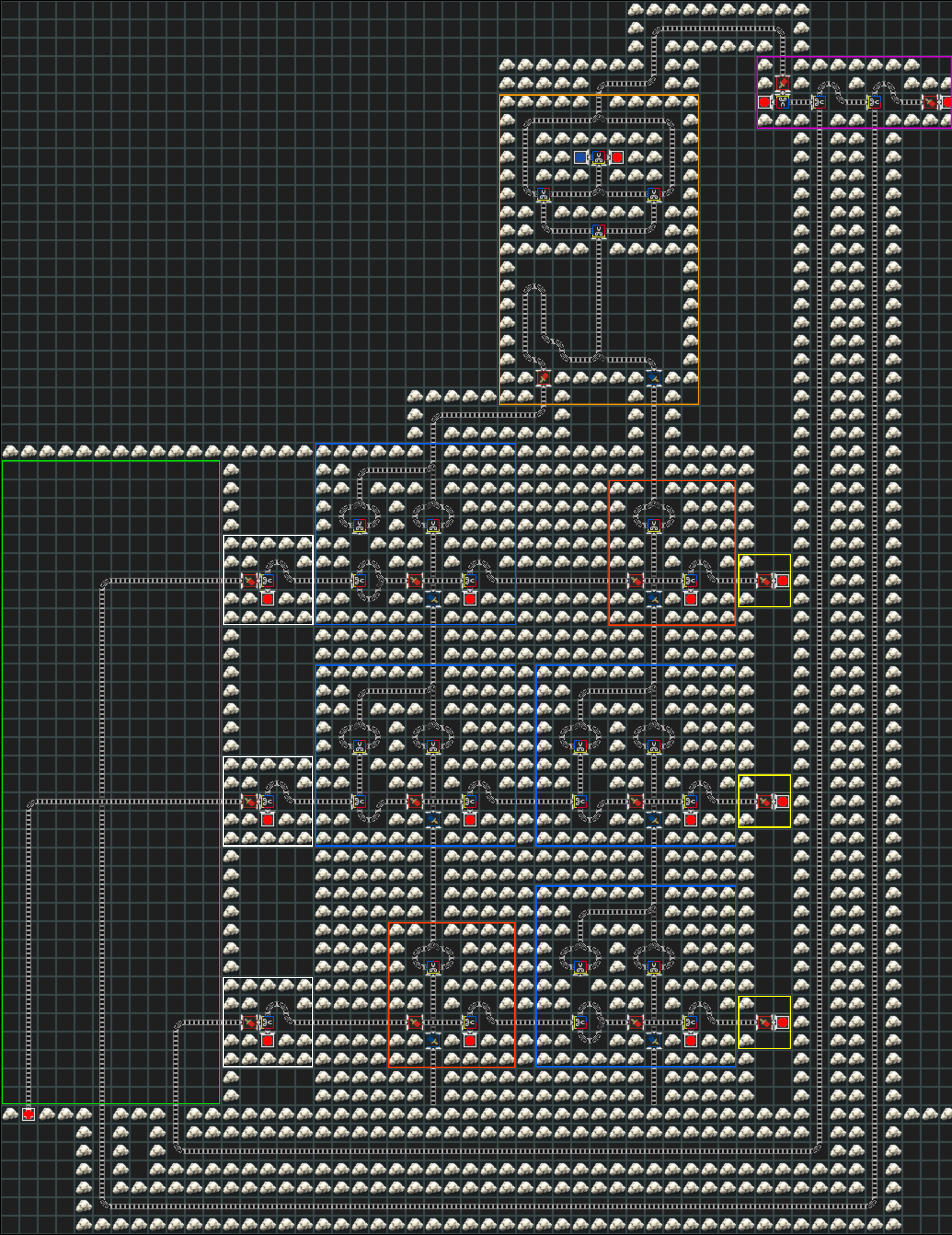}
	\caption{A solved instance of \ty corresponding to the formula $(x_1 \vee x_2) \wedge (x_2 \vee x_3)$. The green area is shown in green and the various gadgets are highlighted in different colors: \gonetimepass in white, \gterminus in yellow, \gignore in red, \gsatisfy in blue, \gand in orange, and \greplicator in purple.}
	\label{fig:instance}
\end{figure}

\end{document}